\documentclass{PoS}

\title{HYDJET++ simulations and reconstruction of the anisotropic flow in Pb+Pb collisions at the LHC}

\ShortTitle{Anisotropic flow with HYDJET++ model}

\author{~L.V.~Bravina, \speaker{G.Kh.~Eyyubova}$^a$$^b$, E.E. Zabrodin$^a$$^b$ \\
       \llap{$^a$} Universitetet i Oslo, Department of Physics, P.O.Box 1048 Blindern N-0316 Oslo, Norway\\
        E-mail: \email{gyulnare@student.matnat.uio.no}}

\author{V.L. Korotkikh,  I.P. Lokhtin, L.V. Malinina, S.V. Petrushanko, A.M. Snigirev\\
        \llap{$^b$} M.V. Lomonosov Moscow State University, D.V. Skobeltsyn
Institute of Nuclear Physics, 119991, Moscow, Russia\\
 E-mail: \email{vlk@lav01.sinp.msu.ru}
}

\abstract{
The azimuthal anisotropy of charged particles in heavy ion collisions is
an important probe of quark-gluon plasma evolution at early stages. In the
 present paper the elliptic flow pattern in Pb+Pb collisions at
$\sqrt{s}=5.5$ \rm{TeV} is analyzed for different hadron species in the
frameworks of HYDJET++ Monte-Carlo model. The influence of resonance decays on particle flow is investigated. The different methods of
elliptic flow reconstruction are compared under LHC conditions.
}

\FullConference{High-pT Physics at LHC -09\\
         February 4- 7 2009\\
         Prague, Czech Republic}

\begin{document}
\section{Introduction}
In non-central collisions between two nuclei the
beam direction and the impact parameter vector define a reaction plane for
each event. The observed particle yield versus azimuthal angle with
respect to the event-by-event reaction plane gives information on
the early collision dynamics \cite{Ollit,Sorge}. An initial nuclear
overlap region has an ``almond'' form at non-zero impact parameter.
If the produced matter interacts and thermalizes, pressure is built
up within the almond shaped region leading to anisotropic pressure
gradients. This pressure pushes against the outside vacuum and the
matter expands collectively. The result is an anisotropic azimuthal angle distribution of the
detected particles. One can expand this azimuthal angle distribution in a
Fourier series. The second coefficient of the expansion $v_2$ is
called  the elliptic flow.

It was found \cite{Sorge, Kolb} that anisotropic flow is self-quenching phenomenon since it reduces
 spatial anisotropy as it evolves. Therefore, observed elliptic flow
must originate at early stages of the collision when the anisotropy is still present in the system.
There is no elliptic flow generated when the spherical symmetry is restored in the system.
Thus the elliptic flow keeps information about hot and dense matter created in relativistic heavy ion collisions.

As the fireball expands, its temperature and energy density drop. Finally, at the freeze-out stage,
the system breaks up into hadrons and their resonances. The effect of resonance decays,
i.e. final state interaction, on the resulting elliptic flow of particles is quite important.
For instance, it can explain partly  the observed deviation of pion elliptic flow from
the so-called constituent quark scaling \cite{Greco}.
 The exact resonance decay kinematics at very low momenta can probably be accounted for the reduction of
$v_2$ coefficient for pions at midrapidity in the hydrodynamical calculation
\cite{Hirano}.

In this work we employ the HYDJET++ model \cite{hydjet++,hydjet++2} to estimate the azimuthal anisotropy of particles
in Pb+Pb collisions
at LHC energy and to study the influence of resonance decays, jet production and jet quenching on elliptic flow,
and also to test the different methods of $v_2$ restoration for different particles.

\section{Simulation of elliptic flow with HYDJET++ model}
The HYDJET++ model \cite{hydjet++,hydjet++2} represents a superposition of soft and hard parts.
These parts are independent and their contribution to the total multiplicity production depends
 on collision energy, centrality and is tuned by model parameters.
The hard part of the model is identical to the hard part of the
HYDJET model \cite{hydjet} and has the possibility to account for
jet quenching effect. The soft part of HYDJET++ event represents the
"thermal" hadronic state where multiplicities are determined under
assumption of thermal equilibrium \cite{fastmc}. Hadrons are produced
on the hypersurface represented by a parametrization of relativistic
hydrodynamics with given freeze-out conditions. Feed down of
hadronic resonances is taken into account. HYDJET++ is capable of
reproducing the bulk properties of heavy ion collisions at RHIC, i.e.
hadron spectra and ratios, radial and elliptic flow, femtoscopic
momentum correlations, as well as high-p$_T$ hadron
spectra~\cite{hydjet++,hydjet++2}.

\begin{figure}
\center
\includegraphics[width=0.7\textwidth]{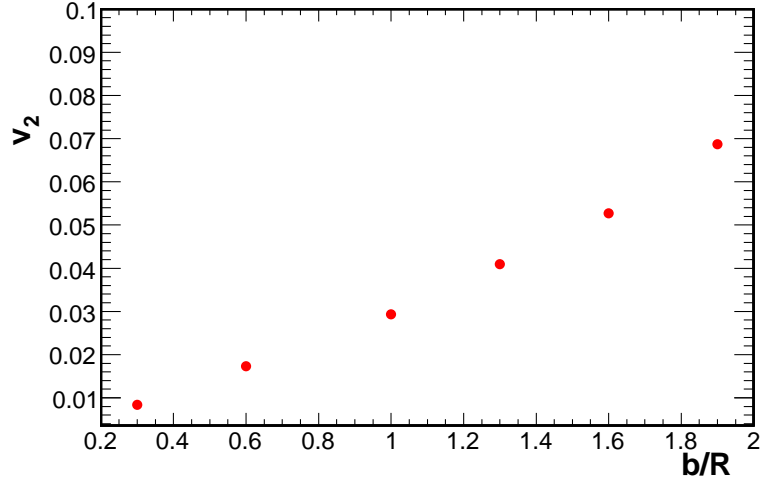}
\caption{The impact parameter dependence of elliptic flow in
HYDJET++  model for Pb+Pb collisions at LHC energy.
\label{v2_b_model} }
\end{figure}

\begin{figure}
\center
\includegraphics[width=0.7\textwidth]{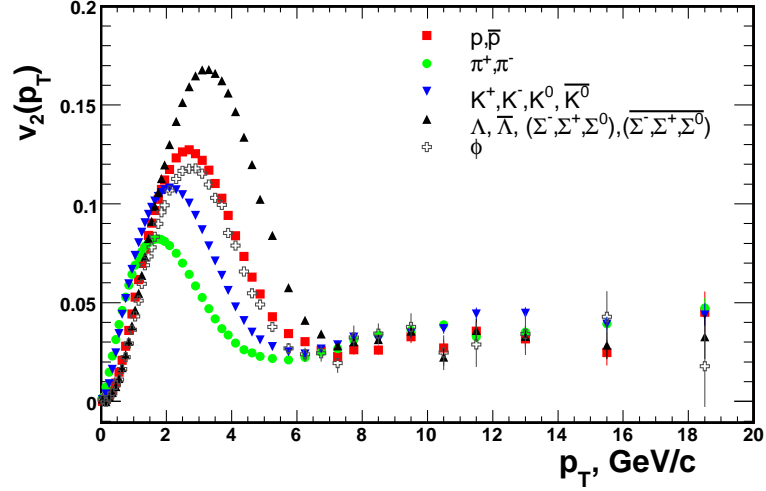}
\caption{ The $p_T$-dependence of elliptic flow in HYDJET++  model
for different hadron species
 produced in Pb+Pb collisions at $\sqrt{s}=5.5$ TeV with centrality 42\%.
\label{v2_pt_id} }
\end{figure}

Fig.~\ref{v2_b_model} presents the impact parameter $b$ dependence of the elliptic flow coefficient
$v_2$ of charged particles. The coefficient
$v_2$ is defined here as the cosine of twice the azimuthal angle
of a particle relative to the reaction plane angle $\Psi_R$ (which is known
in each simulated event), and averaged over all charged hadrons in
each event, $v_2=\langle \cos(2\varphi-\Psi_R)\rangle$. As expected, the elliptic flow coefficient grows with increasing
impact parameter (i.e. with increasing of azimuthal anisotropy of
initial nuclear overlap region).
Further we consider elliptic flow at fixed centrality,
$\sigma/\sigma_{geo}=42\%$ ($b\approx 1.3R$). Fig.~\ref{v2_pt_id} shows the $p_T$
dependence of elliptic flow coefficient for most abundant hadrons,
i.e. pions, kaons, protons, lambdas and sigmas.
This behavior of elliptic flow  can be explained in following way.
 The flow of hydro part rises monotonically up to $v_2\simeq 0.5$ at $p_T \simeq 6$ GeV/c while the
relative contribution of hydro part to particle multiplicity decreases with
$p_T$, so the particles with $p_T \gtrsim 6$ GeV/c are produced only
through jets (hard part). The flow of the jet part is close to zero in this
$p_T$ range. It results in initial rise of $v_2(p_T)$ followed by the fall off at $p_T \gtrsim 3$ GeV/c.
The jet part at $p_T \gtrsim 7$ GeV/c
also presents some amount of flow ($\approx 4\%$) at the LHC
energies due to jet quenching effect. The energy loss of the high
$p_T$ partons depends on the passing length of the anisotropic
matter thus giving the different yield of the high-$p_T$ partons in
the in-plane $(x,z)$ and out-of plane $(y,z)$ directions.

The pronounced feature of the RHIC experimental data \cite{PHENIX}
reproduced by HYDJET++ in Fig.~\ref{v2_pt_id} is the crossing of
baryon and meson branches. In HYDJET++ at low $p_T$ the hydro part
dominates and the flow is strictly ordered by particle masses. The
lighter particles (pions, kaons) have the larger flow than heavier (protons, lambdas).

The slope of the
heavy particles $p_T$-spectra is steeper then one of the light particles,
as a result the hydro part dominates till larger $p_T$ values for heavier particles:
i.e. for pions hydro dominates till $\sim 4$ GeV/c, for
protons it dominates till $\sim 5$ GeV/c.
As a result at $p_T>4$ GeV/c the mass ordering
changes on the opposite:
the heaviest particles have the largest flow.

\begin{figure}
\center
\includegraphics[width=0.7\textwidth]{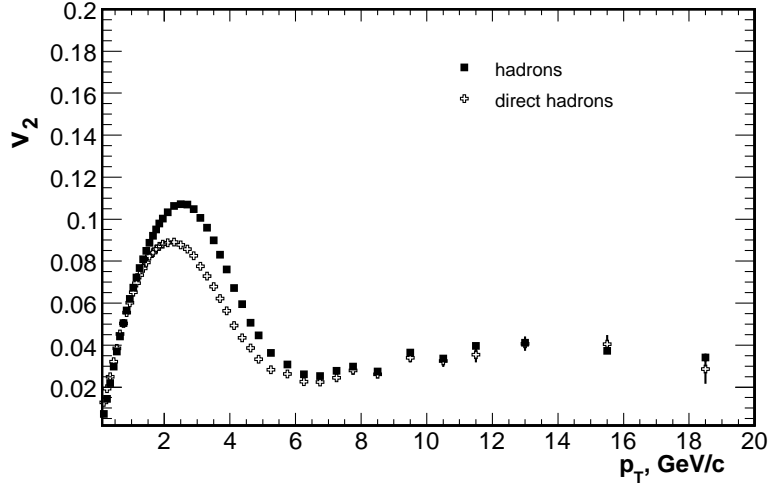}
\caption{The $p_T$-dependence of elliptic flow in HYDJET++ model
 produced in Pb+Pb collisions for all hadrons and
for direct hadrons at $\sqrt{s}=5.5$ TeV with centrality 42\%.
\label{v2_resonance} }
\end{figure}

\begin{figure}
\includegraphics[width=0.5\textwidth]{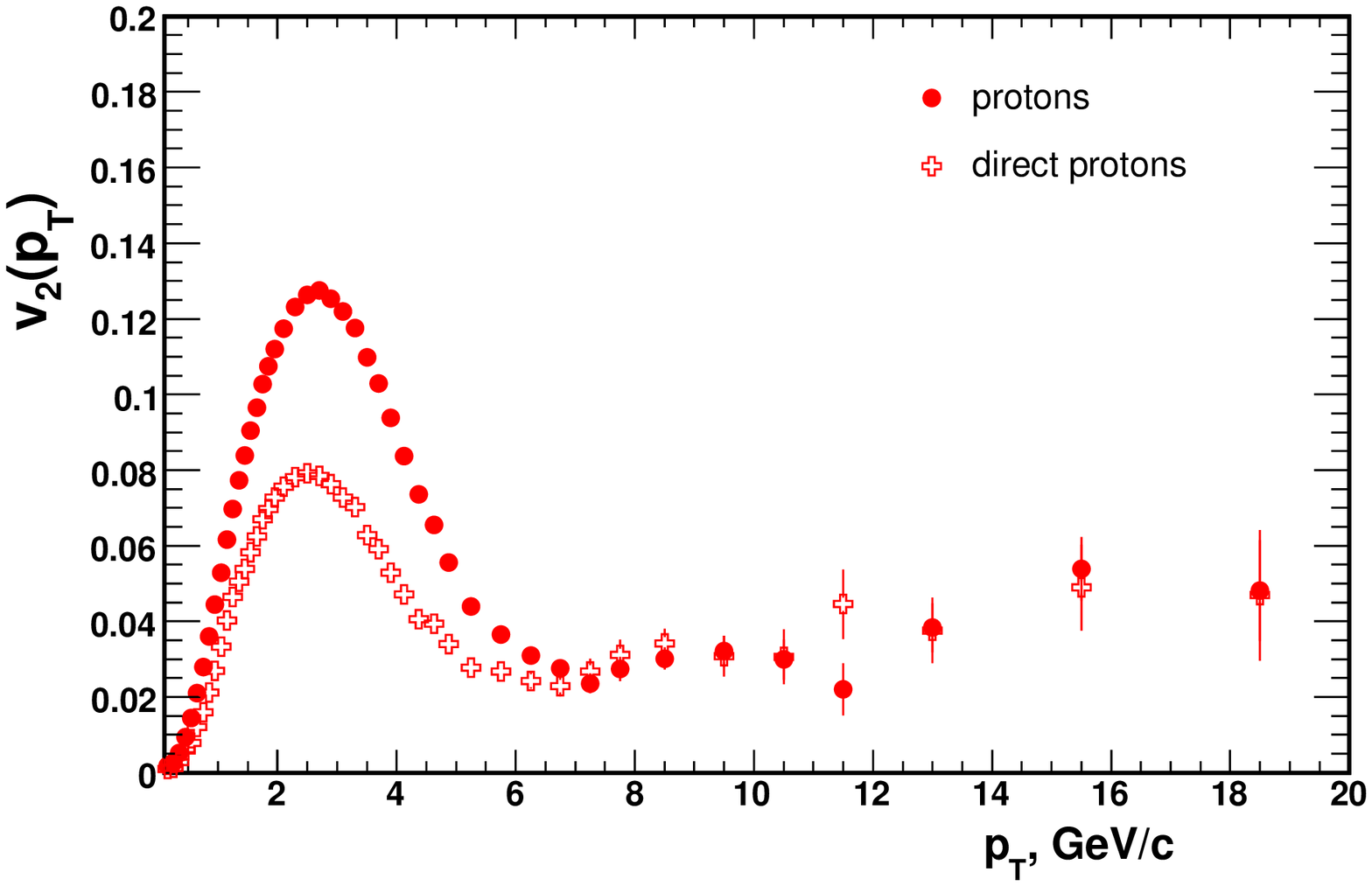}
\includegraphics[width=0.5\textwidth]{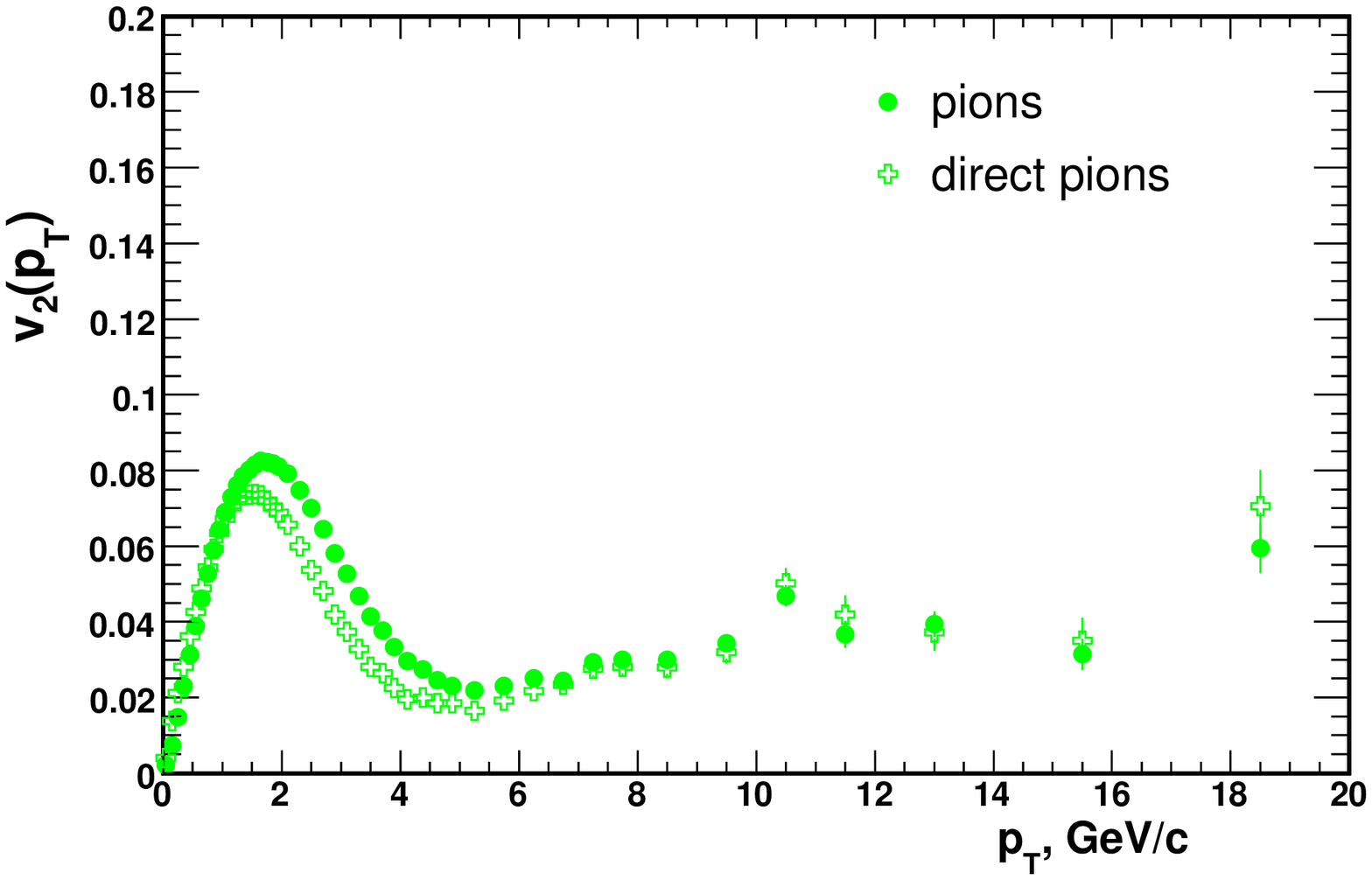}
\includegraphics[width=0.5\textwidth]{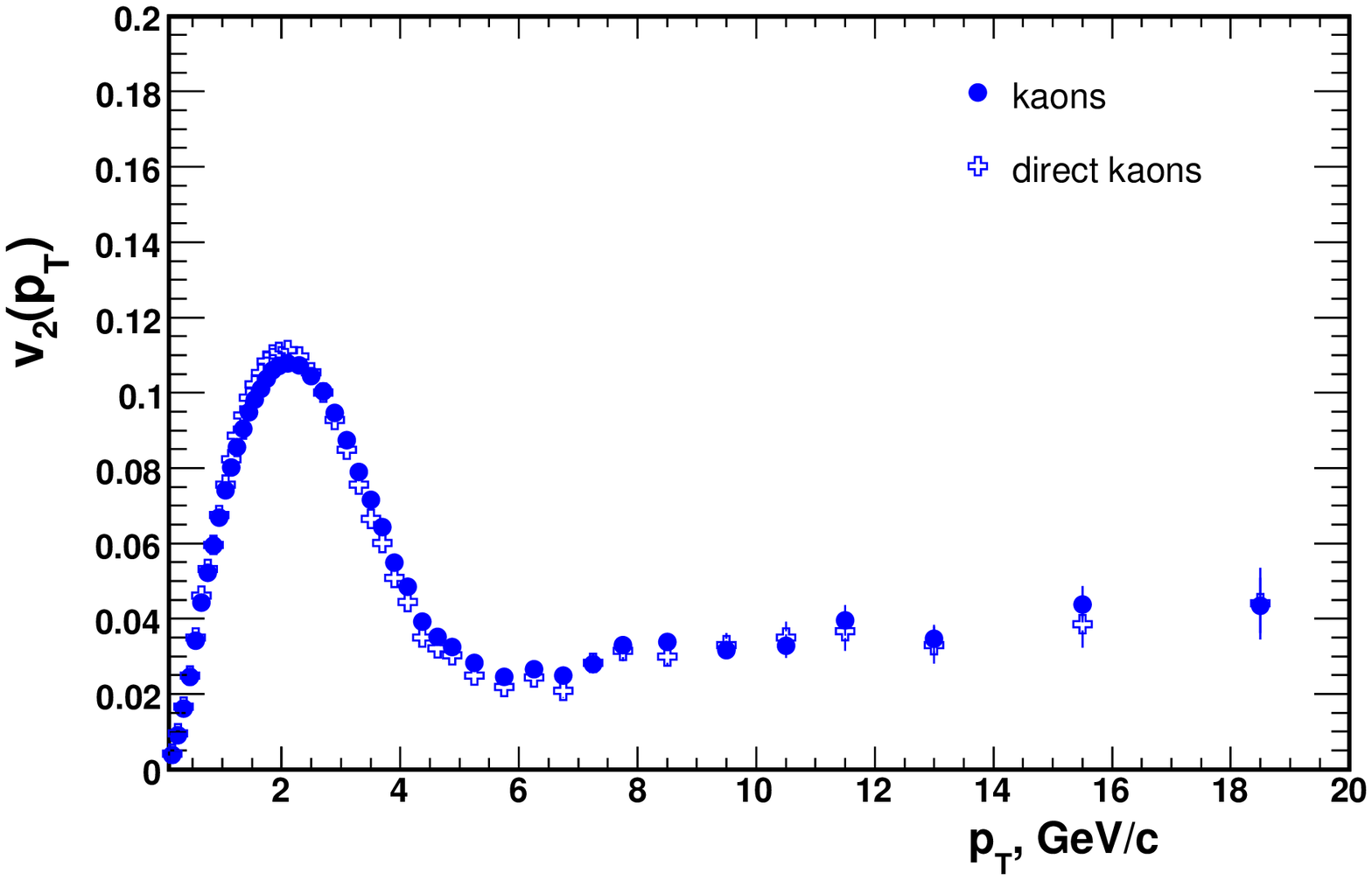}
\includegraphics[width=0.5\textwidth]{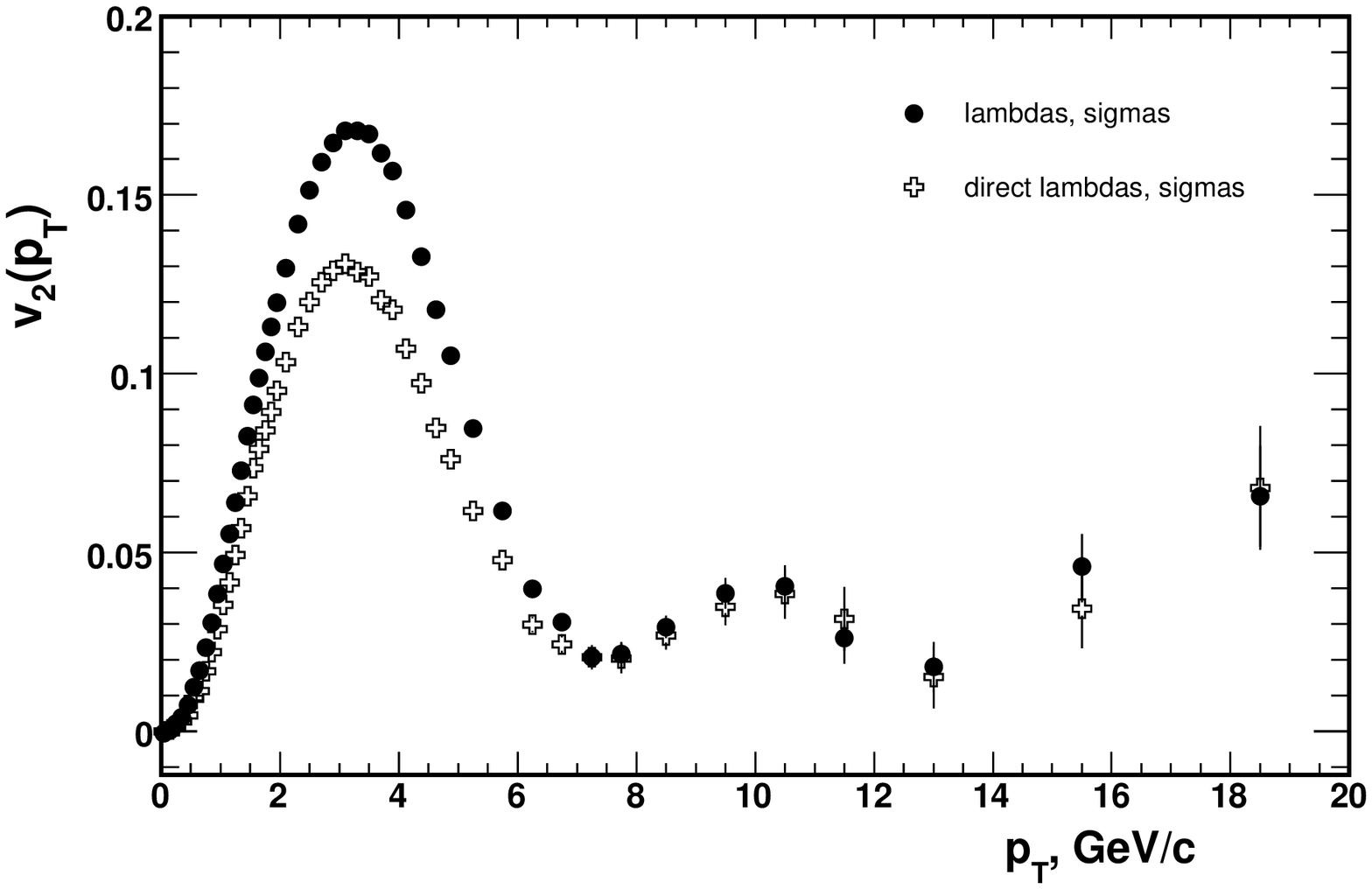}
\caption{The same as Fig. 2, but for protons (upper left), pions
(upper right), kaons (bottom left), lambdas plus sigmas (bottom
right). \label{v2_resonance_id} }
\end{figure}

{\bf The influence of resonance decays.} At RHIC energies the transition from baryon rich matter to meson reach matter was found. 
As was predicted by
Hagedorn, at high energies most of the particles will be
produced through resonance decays with shifting of the average mass to heavier sector. The effect of resonance decays
should be accounted for when one considers the $v_2/n_q$ scaling.
 Table~\ref{tab1} shows the contributions of
direct and resonant production for various hadron species including
feed down from weak decays for our Pb+Pb event sample generated
with HYDJET++ at LHC energies. One can see that 80\% of pions, 70\%
of protons, 61\% of $\Sigma$ and $\Lambda$-hyperons and 56\% of
kaons are produced from resonance decays.
 Figures~\ref{v2_resonance} and \ref{v2_resonance_id} display difference between
$v_2$ of all these hadrons and $v_2$ of only direct hadrons.
\begin{table}
\caption{Yields of the particles produced directly and with resonance decays at midrapidity region,
c=42\%. Weak decays of the strange particles are included.}
    \label{tab1}
    \begin{center}
      \begin{tabular}{|c|c|c|c|c|c|}
\hline
    & $\pi^{\pm}$ & $ K +\bar K $& $p+ \bar p$ & $\Lambda + \bar \Lambda + \Sigma + \bar \Sigma$ & $ \phi $\\
  \hline
all & 860 & 185 & 63.8 & 42.3 & 6.55 \\ \hline
direct & 169 & 81.4 & 18.6 & 14.2  & 6.5 \\\hline

direct \% & 20 \%& 44 \% & 30 \% & 39 \% & 99 \% \\ \hline
      \end{tabular}
    \end{center}
  \end{table}
A degree of the influence of resonance decays on elliptic flow coefficient
is quite different for various hadrons.
The effect is strongest for protons;
rather moderate for $\Sigma$ and $\Lambda$-hyperons and pions; and negligible for kaons.

\begin{figure}
\includegraphics[width=0.5\textwidth]{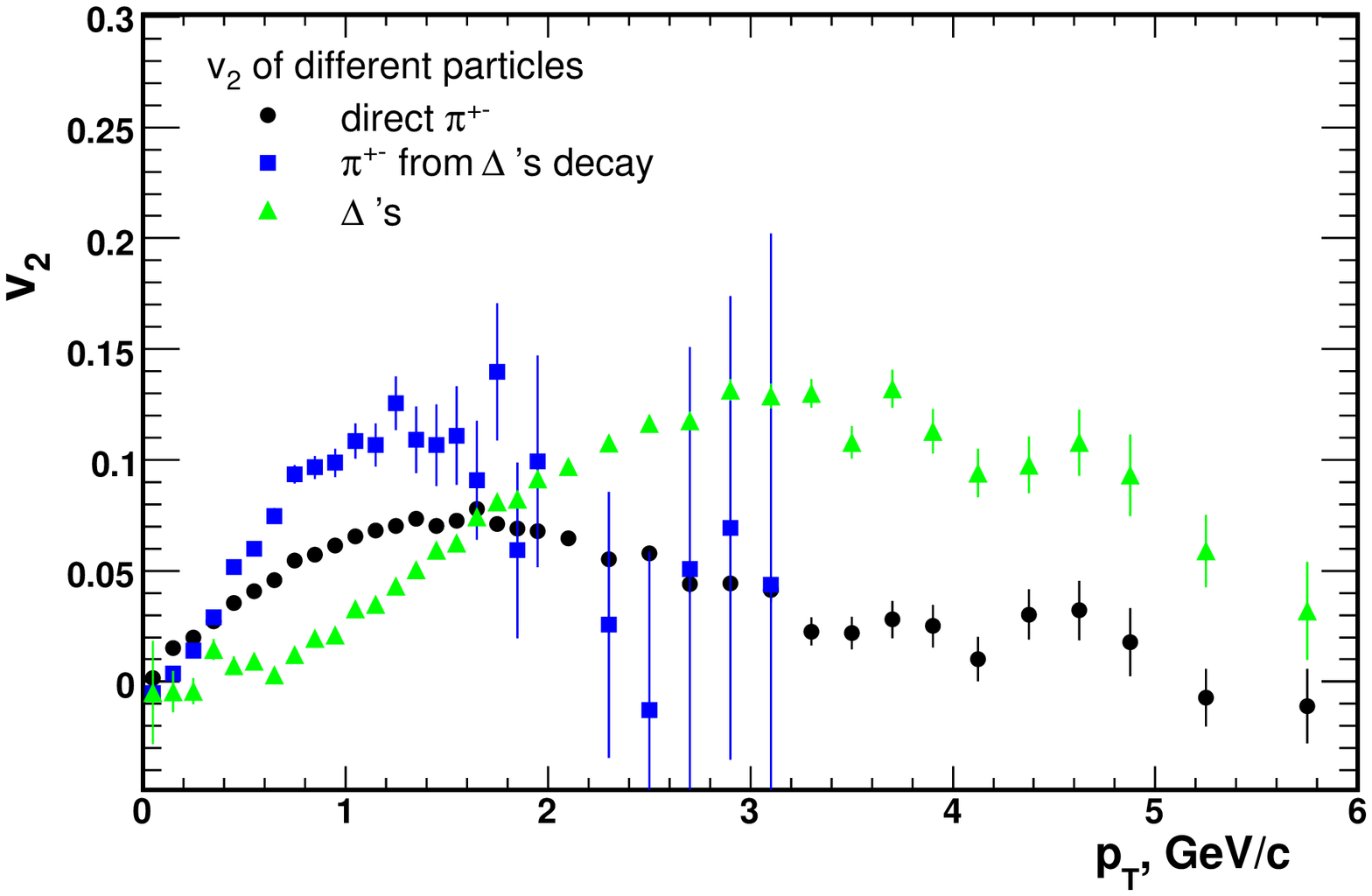}
\includegraphics[width=0.5\textwidth]{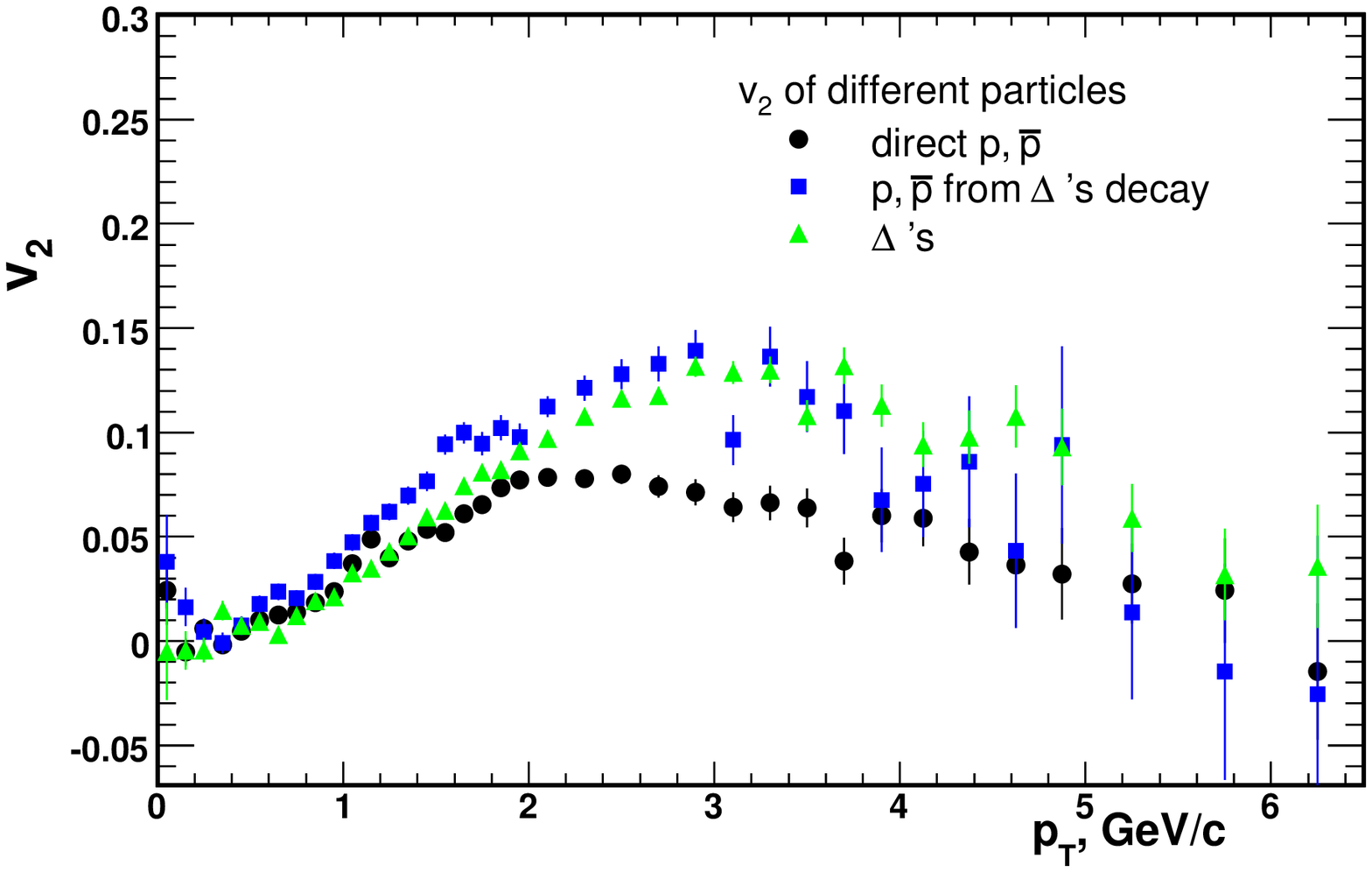}
\includegraphics[width=0.5\textwidth]{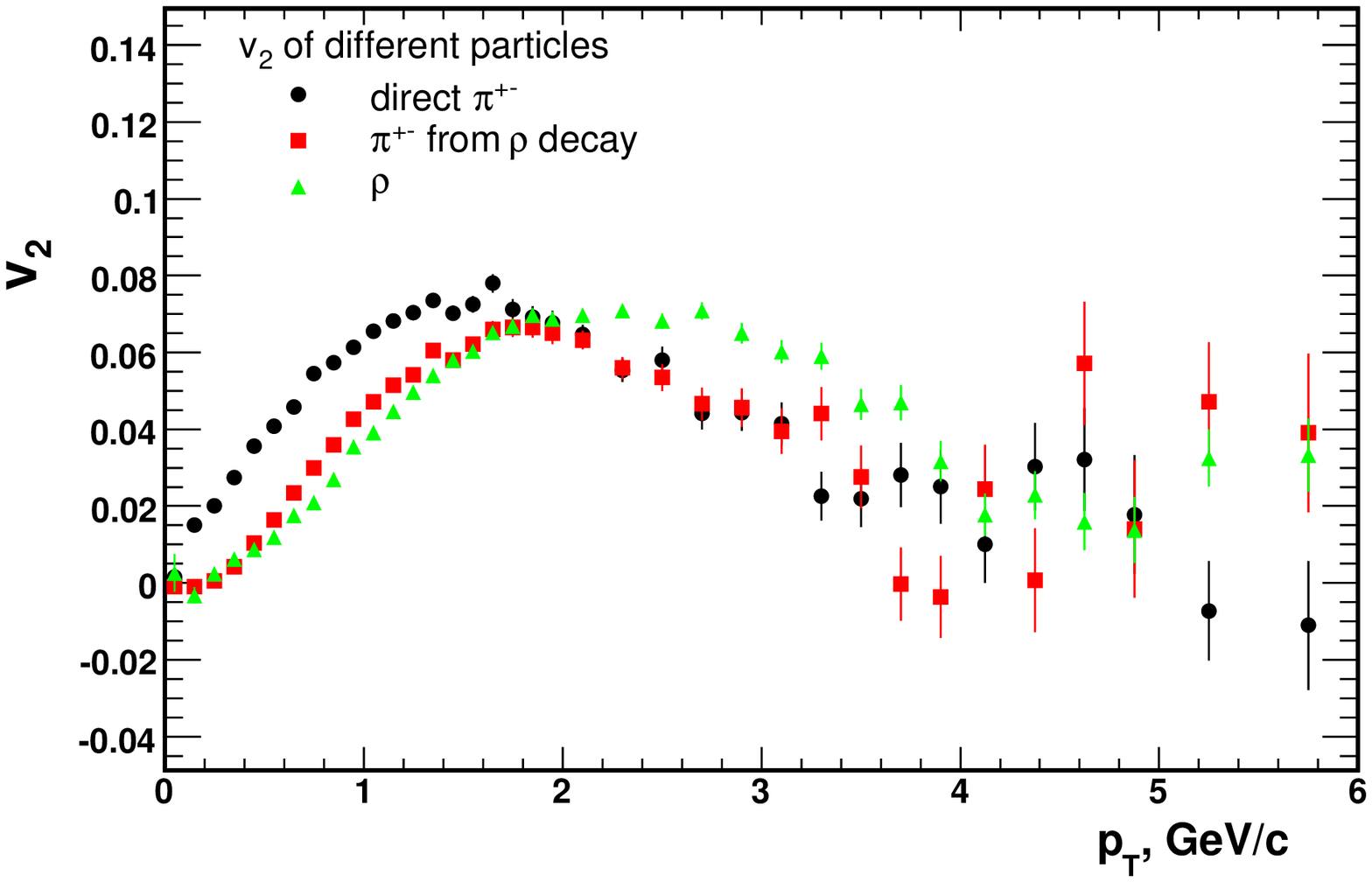}
\includegraphics[width=0.5\textwidth]{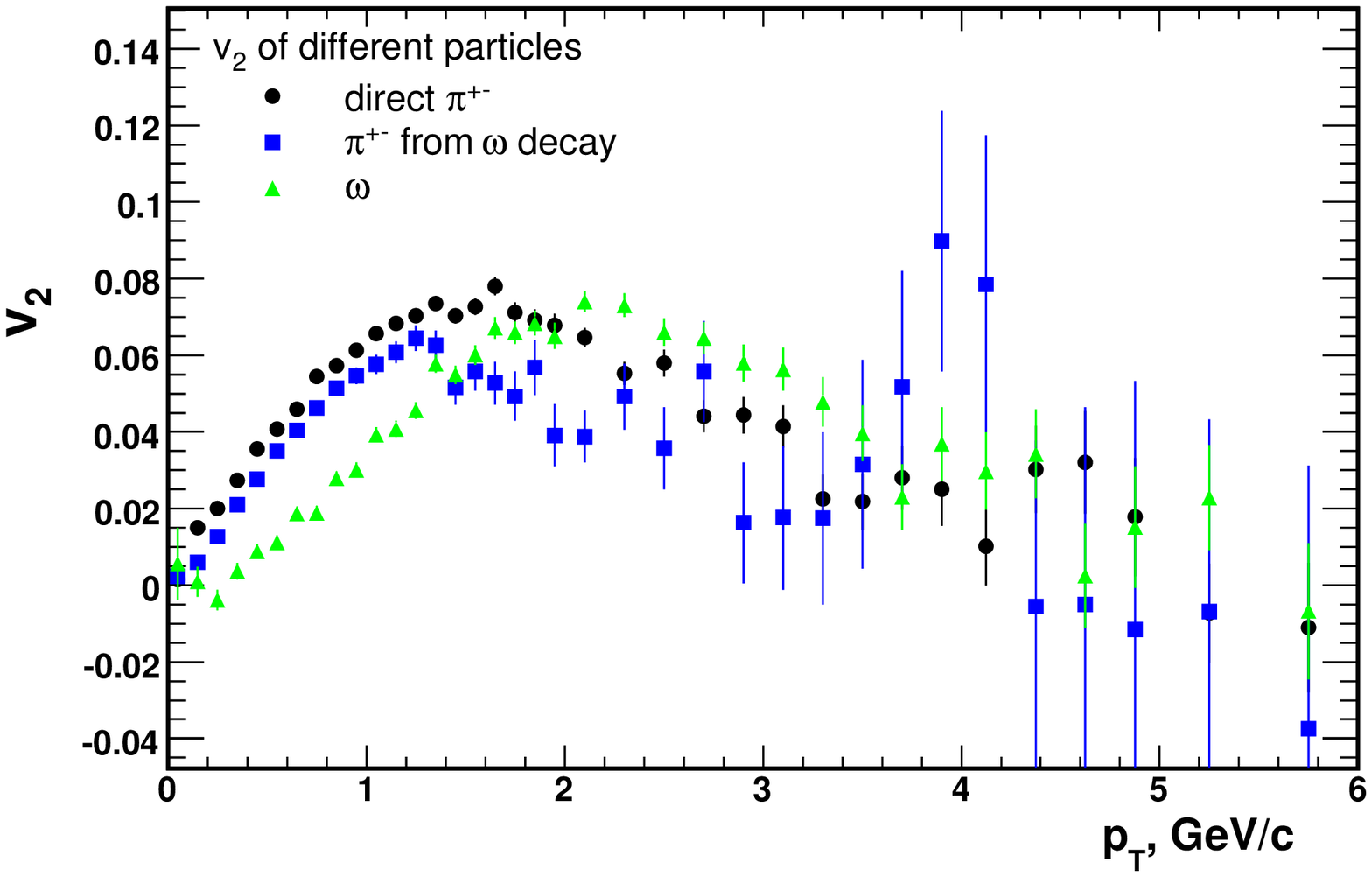}
\caption{ Top: The $p_T$-dependence of elliptic flow for pions (left) and (anti)protons (right)
coming from $\Delta$ decays. Bottom: The elliptic flow $p_T$-dependence of pions coming
from $\rho$ decays (left) and $\omega$ decays (right).
\label{v2_resonance_compare} }
\end{figure}

Let us consider the case of pion and proton flow. The relative
contribution from resonance decays for them is presented in
Table~\ref{tab2}.
\begin{table}
\caption{Yields of the pions and protons produced directly and from resonance decays in HYDJET++, c=42\%.}
    \label{tab2}
    \begin{center}
      \begin{tabular}{|c|c|c|c|c|c|c|}
\hline
    & direct & $\rho$-decay & $ K^{0}$-decay & $\omega$-decay & $\Lambda$-decay & $ \Delta$-decay\\
  \hline
$\pi^{\pm}$ & 22\% & 26\% &16\%  & 11\%& 2.3\% & 1.8\%  \\

 \hline
$p,\bar p$ &30\% &- &- &- &27\%  & 15\%\\
\hline
  \end{tabular}
    \end{center}
  \end{table}
Fig.~\ref{v2_resonance_compare} shows differences in secondary pions
and protons from (anti)deltas ($\Delta^{++}, \Delta^{+}, \Delta^{0},
\Delta^{-}$) decays. Because of the kinematics of decay, when heavy
baryon resonance decays into secondary baryon plus pion, the most
part of its momentum is carried by the baryon while the pion is
produced with low $p_T$. Thus, the pion elliptic flow gets an extra
boost at low $p_T$ from the flow of heavy resonances
(Fig.~\ref{v2_resonance_compare}, top left).
On the other hand, the secondary baryon carries practically the same
$v_2(p_T)$ as mother particle (Fig.~\ref{v2_resonance_compare}, top right).
Note, that most of secondary pions are produced from $\rho$- and
$\omega$-mesons. But momentum distribution for $\rho$-meson and for
secondary pions are quite close, thus the $v_2(p_T)$ almost
coincides for them (Fig \ref{v2_resonance_compare}, bottom left).
The influence of heavy resonance decays on pions flow at $p_T
\lesssim 2$ GeV/c is effectively compensated by influence of
$\rho$-decays. For higher $p_T$ the contribution from $K$-decay determines the
observed excess of the pion flow over direct pion flow.

The secondary protons come from $\Lambda$ and $\Delta$ decays in
approximately equal proportions. They predominantly possess the flow
of these resonances, as can be seen from top left Fig.~\ref{v2_resonance_id}.

One can see from Fig.~\ref{v2_resonance_compare} that hadrons
produced from resonance decays carry the same amplitude of $v_2$ as
that of the mother particle but the maximum may be shifted to the
soft $p_T$ region. Elliptic flow of pions from the $\rho \to \pi\pi$
decay almost coincides with $v_2^{\rho}$ (Fig.~\ref{v2_resonance_compare}, bottom left)
 while in 3-particle decay
$\omega \to \pi\pi\pi$ pions are getting obviously softer $p_T$
distribution, thus their elliptic flow is transferred to the softer
$p_T$ region compared with $v_2^{\omega}(p_T)$ 
(Fig.~\ref{v2_resonance_compare}, bottom right).

Therefore, contributions from resonances sometimes increase and sometimes decrease initial elliptic flow of
 directly produced pions. This effect, especially pronounced at LHC energy, can lead to
violation of the mass-hierarchy in the $v_2(p_T)$ sector.
\section{Reconstruction of elliptic flow at the LHC}
There exists a wealth of anisotropic flow measurement methods,
each having its own advantages and limitations.
Here we apply three wide-spread methods to calculate the
$v_2$ coefficient. One of them uses the event plane angle
determination, and others are cumulant and Lee-Yang zero methods.

The event plane angle, $\Psi_n$, can be
determined from the measured $n$-th harmonics via the standard
method~\cite{Voloshin,voloshin_2}:
 \begin{equation}
\label{f.1} \tan
n\Psi_n=\frac{\sum\limits_{i}w_i\sin(n\varphi_i)}{\sum\limits_{i}w_i
\cos(n\varphi_i)}, ~~~~~~~n\ge1, ~~~0\le\Psi_n<2\pi/n,
\end{equation}
where $\varphi_i$ is the azimuthal angle of the $i$-th particle and
$w_i$ is the weight. The sum runs over all particles in given event.
The observed value of $v_2^{\rm{obs}}$ is calculated using the event
plane (EP) method by the formula:
\begin{equation}
\label{f.obs_1} v_2^{\rm{obs}}\{EP\}=\langle \cos 2 (\varphi - \Psi_2)\rangle ,
\end{equation}
where event plane angle $\Psi_2$ is the estimate of the true
reaction plane angle $\Psi_R$, the mean was taken over all charged particles in
a given event and then over all events.
Usually the true elliptic flow coefficient is evaluated by dividing
$v_2^{\rm{obs}}$ by the factor $R$
\cite{voloshin_2}, which accounts for the event plane resolution:
\begin{equation}
\label{f.R} v_2\{EP\}=\frac{v_2^{\rm{obs}} \{EP\} }{R}=
\frac{v_2^{\rm{obs}} \{EP\} }{\langle \cos 2 (\Psi_2 - \Psi_R)\rangle} .
\end{equation}
This procedure relies on the assumption that there are no non-flow
correlations (e.g., correlations due to momentum conservation, quantum
statistics, resonance decays, jet production) or that
they are negligible, and also that the full event multiplicity is
large enough. Generally such assumptions are not true. Studying
these effects in real data is a separate and non-trivial task.

In order to avoid the trivial autocorrelation of particles the
event plane angle $\Psi_2$ and hence $R$ are calculated in one angular
distribution sample of event, and $v_2$ in another sample with the
same multiplicity. The samples may be selected in two regions of
pseudorapidity $\eta < 0$ and $\eta > 0$.

The basic idea of the cumulant method is that the
$v_2$ coefficient can be expressed in terms of particle azimuthal
correlations~\cite{wang91,Borg}. The procedure is to construct two-particle correlator or cumulant
\begin{eqnarray}
 \nonumber v_2\{2\}^2= \langle \cos2(\varphi_i-\varphi_j)\rangle
=\langle \cos2((\varphi_i-\Psi_R)-(\varphi_j-\Psi_R))\rangle \\
\simeq \langle \cos2((\varphi_i-\Psi_R)\rangle\langle(\varphi_j-\Psi_R))\rangle.
\label{f.10a}
\end{eqnarray}
As in the first method it is necessary to exclude the autocorrelations.
The approximative equation in the last string of Eq. (\ref{f.10a})
means that the non-flow correlations are small.

The two-cumulant method can be extended to the case of many-particle
correlations \cite{Borg}. It is known as the higher order cumulant
method or the method with Lee-Yang zeroes. The higher order cumulant
methods are less sensitive to non-flow effects.
The Lee-Yang zeroes method \cite{L_Y_Z} refers to the Generation
function as a complex function of variable $r$:
\begin{equation}
\label{f.gen_func} G^{\theta}(ir)= \langle e^{irQ^{\theta}}\rangle,
\end{equation}
where $Q^{\theta}$ is the flow vector and $\theta$ is arbitrary angle.
\begin{equation}
\label{f.flow_vector} Q^{\theta}=\sum_{j}^M\cos2(\varphi_j-\theta).
\end{equation}
The integral $v_2$ value is connected with the first minimum of
module of the Generation function $|G^{\theta}(ir)|$,
\begin{equation}
v_2^{\theta}\{\infty\}\equiv \frac{j_{01}}{M r^{\theta}_0},
\end{equation}
 and the differential value is given by more complex
expression:
\begin{equation}
\label{f.v2_dif} \frac{v_2^{~\theta}(p_T)\{\infty\}}{M
v_2\{\infty\}}\equiv Re\left(\frac{\langle\cos 2(\varphi-
\theta)e^{ir^{\theta}_0 Q^{\theta}}\rangle} {\langle
Q^{\theta}e^{ir^{\theta}_0 Q^{\theta}} \rangle} \right).
\end{equation}

\begin{figure}
\includegraphics[width=0.5\textwidth]{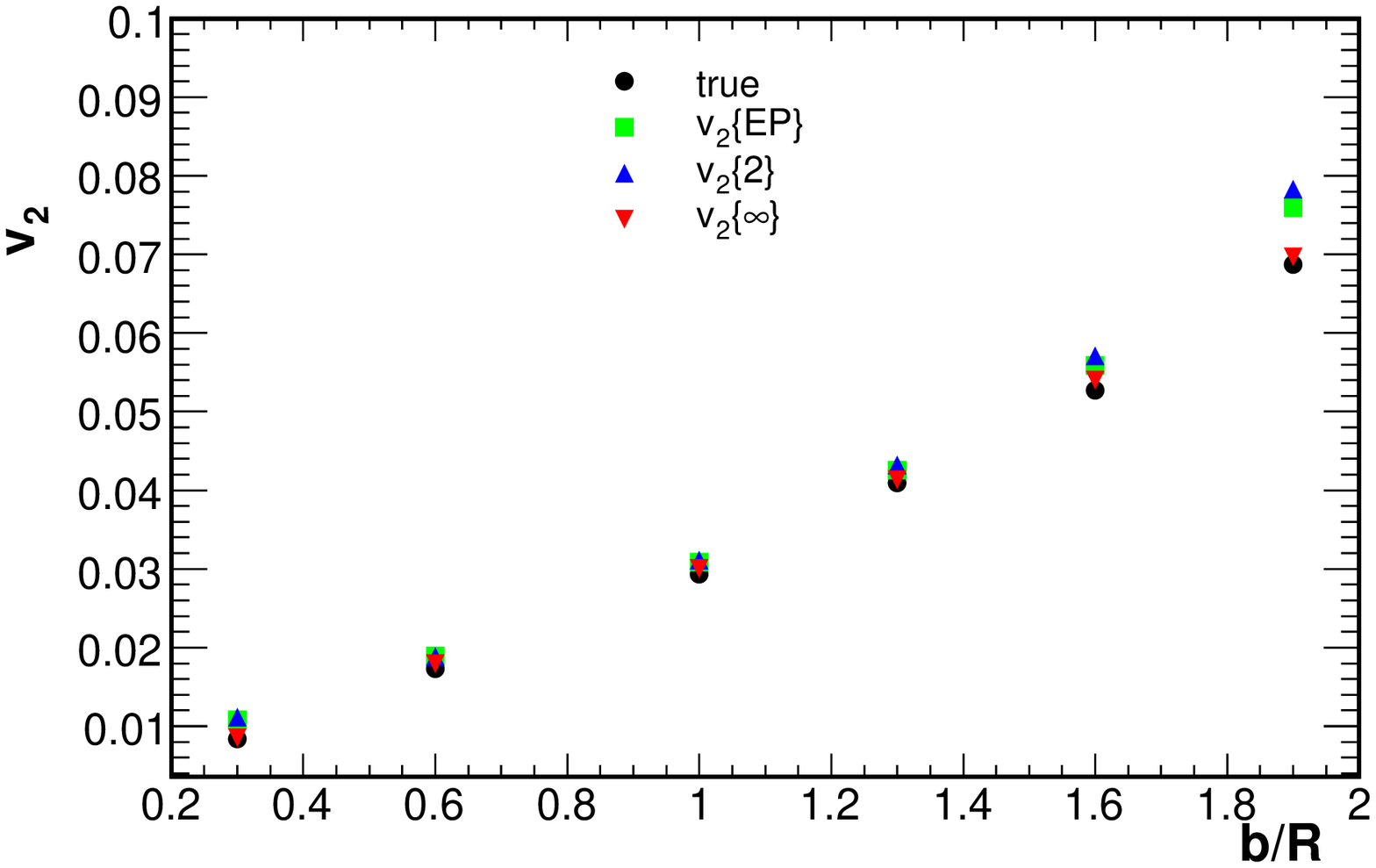}
\includegraphics[width=0.5\textwidth]{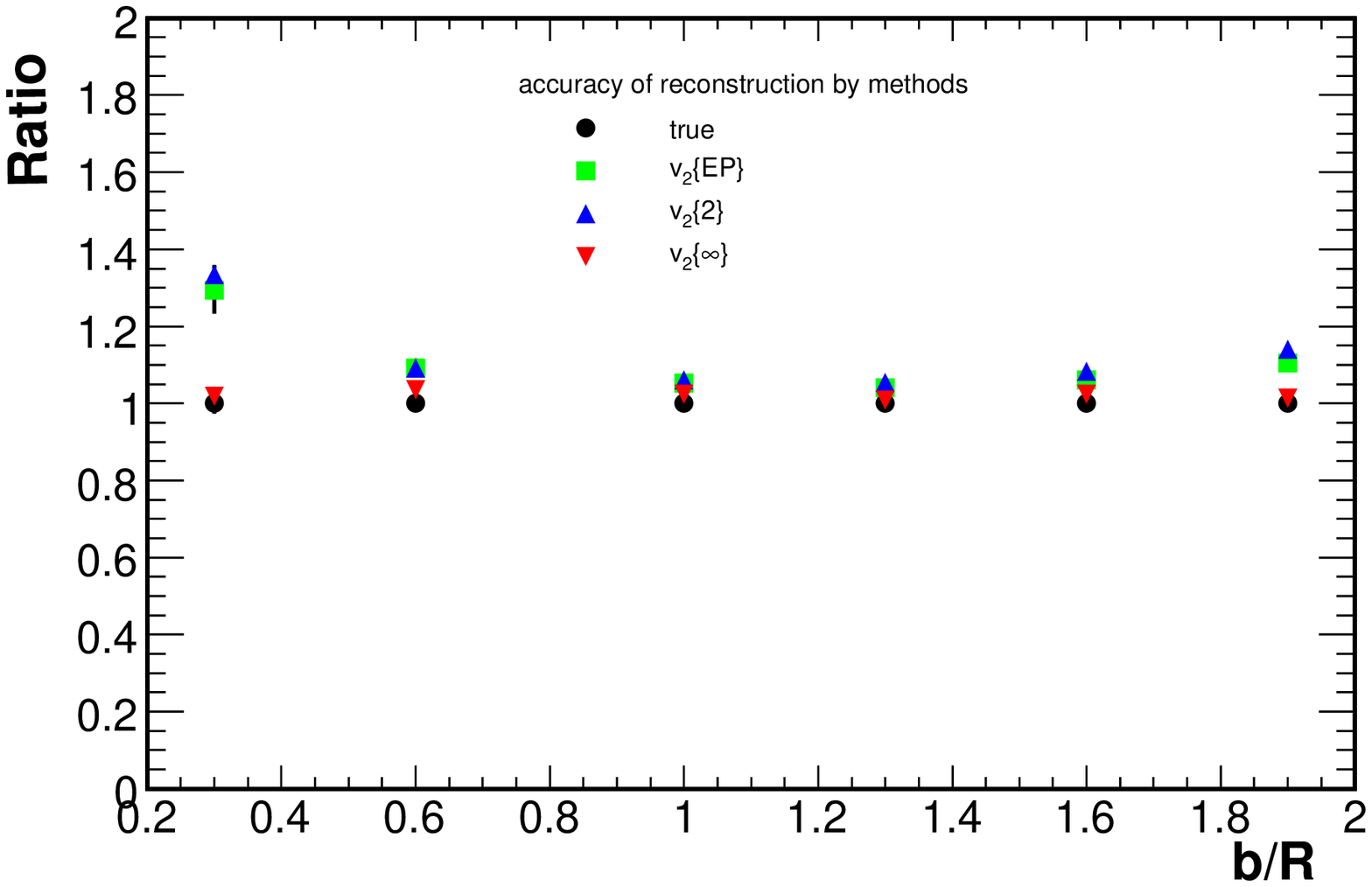}
\caption{(a) The $b$-dependence of elliptic flow in HYDJET++ model
for Pb+Pb at LHC energies reconstructed by different methods.(b) The
ratio of the methods to the true value. \label{v2_methods} }
\end{figure}

The natural resolution parameter that appears in these methods is
$\chi \sim v_2\sqrt{M}$, i.e. methods depend both on strength of the
flow $v_2$ and on multiplicity $M$. The flow increases with rise of
the impact parameter while the multiplicity decreases. The best
reliability of the methods is achieved at midcentral collision (Fig.
\ref{v2_methods}). The overestimation of the true elliptic flow by
$v_2\{EP\}$ and $v_2\{2\}$ methods in most central and peripheral
collision is up to 30\%. This is due to nonflow correlations. We see
that the Lee-Yang zeroes method is good enough to reconstruct $v_2$
in large centrality range except very peripheral collisions where
the multiplicity is too small.

\begin{figure}
\includegraphics[width=0.5\textwidth]{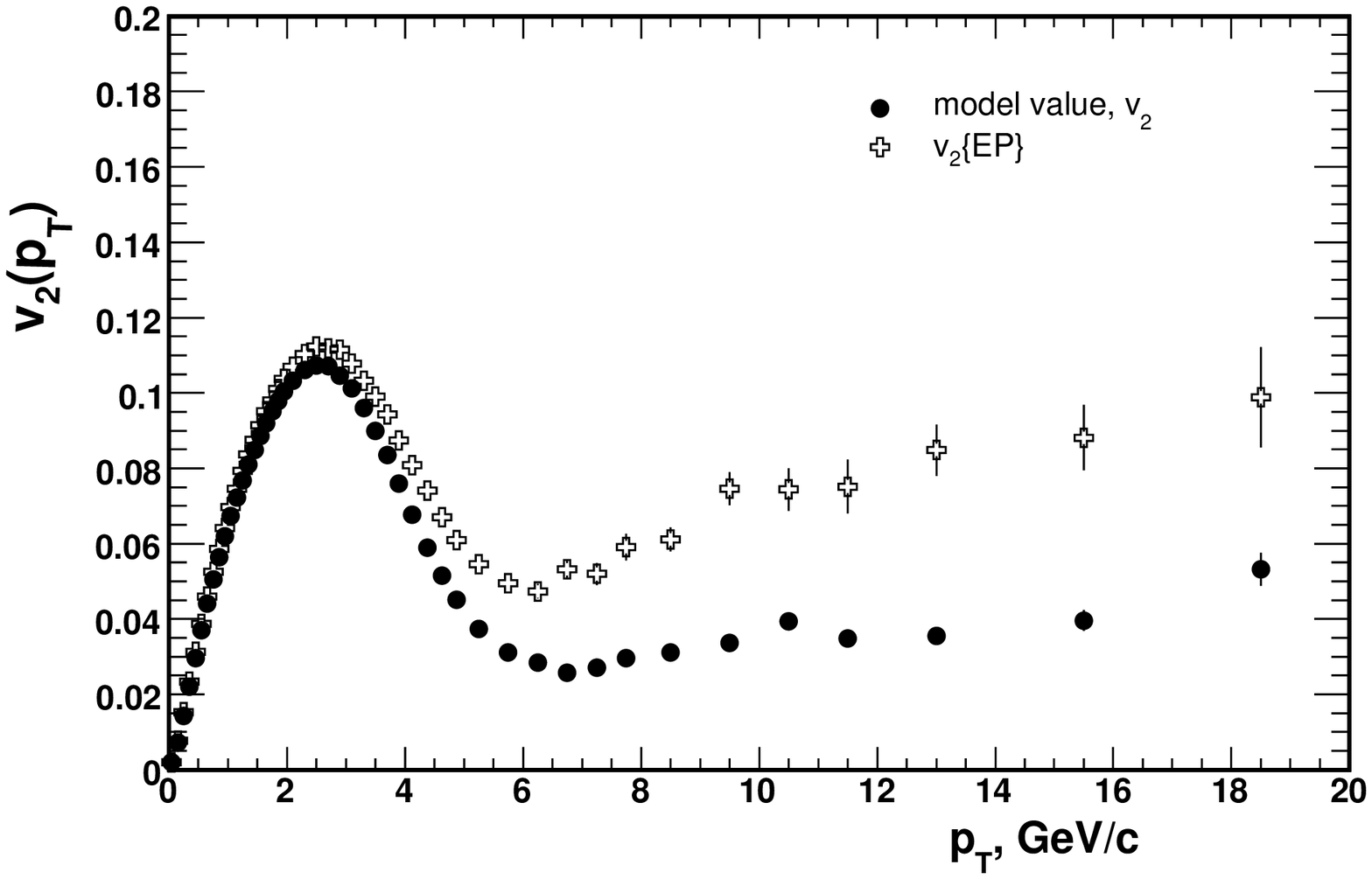}
\includegraphics[width=0.5\textwidth]{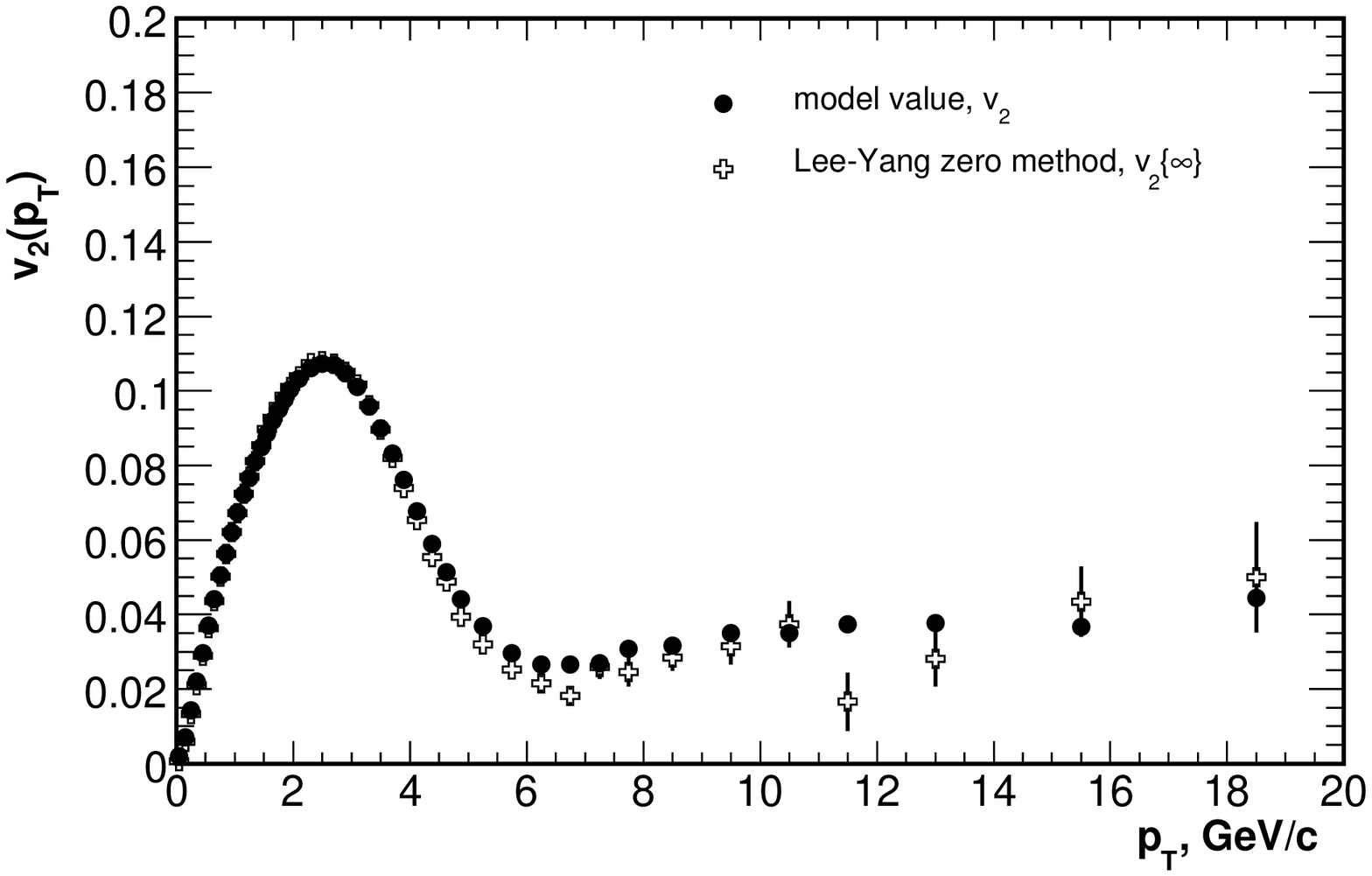}
\caption{The $p_T$-dependence of elliptic flow in HYDJET++ model for
\rm{Pb+Pb} at LHC energies reconstructed by Event plane method
(left) and Lee-Yang zeroes method (right). Centrality is 42\%,
charged hadrons. \label{v2_method_pt} }
\end{figure}

The origin of nonflow correlation is mostly due to jets. It can be
revealed from Fig.~\ref{v2_method_pt}
 where the event plane method severely overestimates the original elliptic flow at high $p_T$.
The possibility of $v_2$ reconstruction for different hadron species
based on the generalization of Lee-Yang zeroes
method~\cite{lyz1,lyz2} at the LHC is under investigation.
\section{Conclusion}
The elliptic flow pattern in Pb+Pb collisions at $\sqrt{s}=5.5$
\rm{TeV} is analyzed for different hadron species in the frameworks
of HYDJET++ Monte-Carlo model. Resonance decays and in-medium jet
fragmentation result in the smearing of hydro-induced mass-ordering
of elliptic flow coefficients $v_2$ for different hadron species in
low- and high-$p_T$ domains, respectively. Increase of $v_2$ due to
resonant production is strongest for protons, moderate for $\Sigma$
and $\Lambda$-hyperons and pions, and negligible for kaons. The
total effect on a given particle specie may contain two different
contributions, either increasing or decreasing the direct flow
$v_2(p_T)$, as we saw on pion example.

 The comparison between three different methods of elliptic flow
 reconstruction under LHC conditions has been performed. The event plane,
the two-particle correlation and Lee-Yang zeroes methods show the
different restoration power. The event plane and two-particle
correlation methods work well for low-$p_T$ region where jet
influence is negligible, while Lee-Yang zeroes method is able to
remove non-flow (jet) correlations at high $p_T$.


 This work was supported, in part, by the QUOTA Program, Norwegian Research Council (NFR) under contract No 185664/V30,
 Russian Foundation for Basic Research (grants No 08-02-91001 and No
08-02-92496), Grants of President of Russian
 Federation (No 107.2008.2 and No 1456.2008.2) and Dynasty Foundation.


\begin{thebibliography}{99}

\bibitem{Ollit} J.-Y. Ollitrault, \emph{Phys. Rev.} {\bf D 46}, (1992) 229.
\bibitem{Sorge} H. Sorge, \emph{Phys. Rev. Lett.} {\bf 82}, (1999)  2048.

\bibitem{Kolb}  P.~F.~Kolb, U.~W.~Heinz, \emph{Quark qluon Plasma 3}, World Scintific, Singapore 2003 
[arXiv:nucl-th/0305084].


\bibitem{Greco}
  V.~Greco and C.~M.~Ko,
  \emph{Phys. Rev.}  {\bf C 70} (2004) 024901.

\bibitem{Hirano}
  T.~Hirano,
 \emph{ Phys. Rev. Lett.}  {\bf 86} (2001) 2754.
\bibitem{hydjet++} I.P.~Lokhtin, L.V.~Malinina, S.V.~Petrushanko,
A.M.~Snigirev, I.~Arsene and K.~Tywoniuk, \emph{Comput. Phys. Commun.}
{\bf 180} (2009) 779.

\bibitem{hydjet++2} I.P.~Lokhtin et al., in this Proceedings,
{\tt arXiv:0903.0525}.

\bibitem{hydjet} I.P.Lokhtin and A.M.Snigirev,
 \emph{Eur. Phys. J.} {\bf C 46},(2006) 211.

\bibitem{fastmc} N.S. Amelin et.al, \emph{Phys. Rev.} {\bf C 74}, (2006) 064901;
\emph{Phys. Rev.} {\bf C 77}, (2008) 014903.

\bibitem{PHENIX}
  A.~Adare et al. [PHENIX Collaboration],
  \emph{Phys.\ Rev.\ Lett.}  {\bf 98},  (2007) 162301.


\bibitem{Voloshin} S.~A.~Voloshin and Y.~Zhang, \emph{Z. Phys.} {\bf C 70},(1996) 665.
\bibitem{voloshin_2}  A.~M.~Poskanzer and S.~A.~Voloshin,
  \emph{Phys. Rev.}  {\bf C 58},(1998) 1671.
\bibitem{wang91} S. Wang et al.,\emph{ Phys. Rev.} {\bf C 44} (1991) 1091 .
\bibitem{Borg} N. Borghini, P.M. Dinh and J.-Y. Ollitrault,
  \emph{Phys.\ Rev.} {\bf C 64}, (2001) 054901.
\bibitem{L_Y_Z}R. S. Bhalerao, N. Borghini, J.-Y. Ollitrault,  \emph{ Nucl. Phys.} {\bf A 727}, (2003) 373.

\bibitem{lyz1} N.~Borghini, R.S.~Bhalerao, J.-Y.~Ollitrault,
 {\emph J.Phys. G} {\bf 30} (2004) S1213.

 \bibitem{lyz2} B.I.~Abelev et al. [STAR Collaboration], \emph{Phys. Rev.}
 {\bf C 77} (2008)  054901.

\end{thebibliography}
\end{document}